\documentclass[aps,byrevtex,nofootinbib]{revtex4}
\usepackage{amsmath,amssymb}
\usepackage{graphicx}
\usepackage{color}
\usepackage{comment}
\usepackage{bbold}

\DeclareMathOperator{\Tr}{tr}
\DeclareMathOperator{\sTr}{str}

\begin{document}

\unitlength=1mm
\def\a{{\alpha}}
\def\b{{\beta}}
\def\d{{\delta}}
\def\D{{\Delta}}
\def\e{{\epsilon}}
\def\g{{\gamma}}
\def\G{{\Gamma}}
\def\k{{\kappa}}
\def\l{{\lambda}}
\def\L{{\Lambda}}
\def\m{{\mu}}
\def\n{{\nu}}
\def\w{{\omega}}
\def\O{{\Omega}}
\def\S{{\Sigma}}
\def\s{{\sigma}}
\def\t{{\tau}}
\def\th{{\theta}}
\def\x{{\xi}}

\def\ol#1{{\overline{#1}}}

\def\Dslash{D\hskip-0.65em /}
\def\dslash{{\partial\hskip-0.5em /}}
\def\vslash{{\rlap \slash v}}
\def\qbar{{\overline q}}

\def\CPT{{$\chi$PT}}
\def\QCPT{{Q$\chi$PT}}
\def\PQCPT{{PQ$\chi$PT}}
\def\tr{\text{tr}}
\def\str{\text{str}}
\def\diag{\text{diag}}
\def\order{{\mathcal O}}
\def\vit{{\it v}}
\def\vD{\vit\cdot D}
\def\am{\alpha_M}
\def\bm{\beta_M}
\def\gm{\gamma_M}
\def\smb{\sigma_M}
\def\smt{\overline{\sigma}_M}
\def\tb{{\tilde b}}

\def\mc#1{{\mathcal #1}}

\def\Bbar{\overline{B}}
\def\Tbar{\overline{T}}
\def\cBbar{\overline{\cal B}}
\def\cTbar{\overline{\cal T}}
\def\pq{(PQ)}

\def\eqref#1{{(\ref{#1})}}

\newcount\hour \newcount\hourminute \newcount\minute 
\hour=\time \divide \hour by 60
\hourminute=\hour \multiply \hourminute by 60
\minute=\time \advance \minute by -\hourminute
\newcommand{\mydate}{\ \today \ - \number\hour :\number\minute}

\title{Strong isospin breaking with twisted mass lattice QCD}

\author{Andr\'{e} Walker-Loud}
\email[]{walkloud@wm.edu}
\affiliation{Department of Physics, College of William and Mary,
	P.O. Box 8795,
	Williamsburg, VA 23187-8795, USA}

\begin{abstract}
In this work we propose a method for including strong isospin breaking in twisted mass lattice calculations, while preserving flavor identification.  We utilize a partially quenched construction in which the sea quarks are given by the standard twisted mass lattice action while the valence quarks  have an additional strong isospin breaking mass term.  This construction allows for a practical use with existing twisted-mass gauge ensembles.  Additionally, we construct the relevant partially quenched twisted mass chiral perturbation theory for both mesons and baryons to $\mc{O}(m_q^2, m_q a, a^2)$.  We provide explicit expressions for the pion, nucleon and delta masses, as well as the corresponding mass splittings, and discuss the resulting errors from including the strong isospin breaking in the valence sector only.  Finally, we demonstrate how the application of this idea can be used, with mild approximations, to determine the values of both the $up$ and $down$ quark masses.
\end{abstract}

\date{\today}
\maketitle


\section{Introduction}
In the last few years, we have witnessed the development of twisted mass lattice QCD~\cite{Frezzotti:1999vv,Frezzotti:2000nk} grow from its initial quenched studies~\cite{Jansen:2003ir,AbdelRehim:2005gz,Jansen:2005gf,Jansen:2005cg,Jansen:2005kk} into a fully viable method for including two flavors of dynamical fermions~\cite{Boucaud:2007uk,Blossier:2007vv,Boucaud:2008xu,Alexandrou:2008tn,Blossier:2009bx}, with more recent extensions to $2+1+1$ flavors~\cite{Chiarappa:2006ae,Baron:2008xa}.  The twisted mass lattice action offers the promise of automatic $\mc{O}(a)$ improvement (where $a$ denotes the lattice spacing), provided one can ``tune to maximal twist"~\cite{Frezzotti:2003ni,Frezzotti:2004wz,Frezzotti:2005gi,Aoki:2004ta,Sharpe:2005rq,Aoki:2006nv}.  Further, the twisted mass term protects the Wilson-Dirac operator from acquiring zero or negative eigenvalues, as the symmetry properties of the twisted mass lattice action prevent additive mass renormalization, thus allowing for a cheaper numerical push to light quark masses.  In a parallel effort, twisted mass chiral perturbation theory was developed for mesons~\cite{Munster:2003ba,Scorzato:2004da,Sharpe:2004ps,Sharpe:2004ny} as well as baryons~\cite{WalkerLoud:2005bt}, both to aid in the understanding of the phase structure of the theory~\cite{Farchioni:2004us} and allow for a combined continuum-chiral extrapolation to the physical point.  For a nice review of twisted mass lattice QCD, see Refs.~\cite{Sint:2007ug,Shindler:2007vp}.

There are a few drawbacks however.  First the action must contain an even number of fermion flavors.  Second, the action at finite lattice spacing breaks parity (although it has a combined parity and flavor transformation symmetry), and thus the study of odd (even) parity states is complicated by their overlap with their lighter even (odd) parity counterparts.  Third, the theory breaks the $SU(2)$ flavor symmetry of light quarks down to $U(1)$.  As an extreme example, this allows for a mixing of the $I=2, I_3=0$ and $I=0$ $\pi\pi$ states~\cite{Buchoff:2008hh}, complicating or rendering impractical the calculation of the $I=0$ ($I_3 = 0$) scattering channels.  Lastly, the inclusion of strong isospin breaking effects ($m_u \neq m_d$) leads to a non-perturbative mixing of the quark flavors~\cite{WalkerLoud:2005bt}, thus complicating the computation of matrix elements in which flavor identification is crucial.  It is this last problem we address in this work.

The inclusion of strong isospin breaking in the twisted mass lattice action will allow for the computation of phenomenologically interesting and important physical matrix elements.  Most notably, it will allow for a lattice determination of the $up$ and $down$ quark masses~\cite{Wilczek:2002wi}.  Further, it will allow for a computation of the isospin violations in the hadron spectrum, for example as was recently done for the neutron-proton mass splitting utilizing a partially quenched framework in which the sea quarks were degenerate~\cite{Beane:2006fk}.

In this work, we propose a partially quenched extension to the standard twisted mass lattice action which introduces a strong isospin violating quark mass in the valence sector, with an average mass equal to the mass of the degenerate pair of twisted mass sea quarks.  After detailing our proposed action in Sec.~\ref{sec:proposal}, we proceed to construct the relevant partially quenched twisted mass chiral perturbation theory for both mesons, to leading order (LO) and next-to-leading order (NLO), as well as that for nucleons and deltas through next-to-next-to leading order (NNLO), Sec.~\ref{sec:PQtmChPT}.  In this section, we also discuss the vacuum alignment of the theory and the hairpin structure from the partial quenching.  We then proceed to determine explicit expressions for the pion, Sec.~\ref{sec:PionMasses} and nucleon and delta masses in Sec.~\ref{sec:baryonMasses}, as well as detail the mass splitting expressions.  Importantly, we show in this partially quenched framework, the NLO corrections to the nucleon (and delta) masses exactly cancel.  Therefore, the expansion of the nucleon mass splittings should be as well behaved as for the pion mass itself.  We then conclude in Sec.~\ref{sec:Discussion}, showing how this method can be used to determine the values of the physical light quark masses.

\section{Proposed Action\label{sec:proposal}}

In order to prove the existence of a positive-definite quark determinant with a twisted-mass lattice action and mass non-degenerate flavors, it was shown the Pauli-matrix used for the isospin breaking must be different from that used with the twisted mass term~\cite{Frezzotti:2003xj}.  However, as was explored in Ref.~\cite{WalkerLoud:2005bt}, with a focus on baryons, because the discretization effects in hadronic quantities are expected to be approximately the same order of magnitude as the light quark effects, this leads to a non-perturbative mixing of the $up$ and $down$ flavors.  For example, with a quark doublet $q = (q_1,q_2)$, an isospin breaking mass term implemented with $\t_3$ would lead to the identification $q_u = q_1$ and $q_d = q_2$ while a twisted-mass term proportional to $\t_1$ would lead to $q_u = (q_1 + q_2)/\sqrt{2}$ and $q_d = (q_1 - q_2)/\sqrt{2}$.  The non-perturbative mixing of these two bases can be resolved, for example, order by order in twisted-mass $\chi$PT by comparing to numerical results.%
\footnote{This is not expected to be as much of an issue in the numerical implementation of the $strange$--$charm$ doublet~\cite{Chiarappa:2006ae,Baron:2008xa}.  This is because the mass splitting term which is used is much bigger than the typical discretization effects, which are typically on the order of the light quark masses.} 

The requirement of different Puali-matrices for the mass splitting and the twisted-mass term arises from the desire to have a positive definite fermion determinant.  This restriction does not apply to the computation of fermion propagators, and naturally leads to the consideration of a partially quenched action.  We begin with the target continuum theory.  We propose an action for which the continuum limit is a partially quenched theory with non-degenerate valence quarks, with an average mass equal to degenerate pair of sea quarks.  In terms of the $valence$-$sea$-$ghost$ quark labels, the quark mass matrix is given by
\begin{equation}
m_Q = \begin{pmatrix}
	m - \d \\
	& m+\d\\
	&&m \\
	&&& m\\
	&&&&m - \d \\
	&&&&& m+\d
	\end{pmatrix}\, ,
\end{equation}
and the continuum Euclidean Lagrangian is given by
\begin{equation}
	\mc{L}_{tm}^{PQ} = \bar{Q} \left[ \Dslash 
	+ m \mathbb{1} 
	-\d \t_3^{v} \right] Q\, ,
\end{equation}
with 
\begin{align}
&Q^T = (u_{val}, d_{val}, u_{sea}, d_{sea}, u_{ghost}, d_{ghost})\, ,&
\\
&\t_3^v = \textrm{diag}(1,-1,0,0,1,-1)\, .&
\end{align}
We know the continuum Lagrangian of the standard implementation of the twisted mass lattice action can be written
\begin{equation}
	\mc{L}_{tm} = \bar{q} e^{i\w \t_3 \g_5 / 2} \left[ \Dslash + m\mathbb{1} \right] e^{i\w \t_3 \g_5 / 2} q
\end{equation}
where $q^T = (u,d)$ and $\t_3$ is the standard Pauli-3 matrix.  Writing this in more standard form
\begin{equation}
	\mc{L}_{tm} = \bar{q^\prime} \left[ \Dslash + m^\prime \mathbb{1} + i \mu \t_3 \g_5 \right] q^\prime\, .
\end{equation}
Therefore, the lattice action we propose can be determined from
\begin{equation}
	\mc{L}_{tm}^{PQ} = \bar{Q} e^{i\w \t_3^{vs} \g_5 / 2}\left[ \Dslash 
	+ m \mathbb{1} 
	-\d \t_3^{v} \right] e^{i\w \t_3^{vs} \g_5 / 2}Q\, ,
\end{equation}
with
\begin{equation}
	\t_{3}^{vs} = \textrm{diag}(1,-1,1,-1,1,-1)\, .
\end{equation}
At maximal twist, $\w = \pi/2$, this Lagrangian is
\begin{equation}
	\mc{L}_{tm}^{PQ} = \bar{Q} \left[ \Dslash 
	+i \mu \t_3^{vs} \g_5
	-i \d \mc{P}_V \g_5
	\right] Q\, ,
\end{equation}
where%
\footnote{Note that $[\t_3^{v}, \t_3^{vs}] = [\t_3^{v}, \mc{P}_V] = [\mc{P}_V, \t_3^{vs}] = 0$.} 
\begin{equation}
	\mc{P}_V = \textrm{diag}(1,1,0,0,1,1)\, ,
\end{equation}
is the valence projector.  This can be thought of a pair of non-degenerate Osterwalder-Seiler valence quarks~\cite{Osterwalder:1977pc}, with masses equal to $m_{OS} = |\mu \pm \d|$.%
\footnote{We thank Steve Sharpe for drawing our attention to this connection.} 

From this point, there are two practical ways one can proceed, which are equivalent in the continuum limit.  Option one is to add the following mass term during the calculation of the valence propagators;
\begin{equation}\label{eq:option1}
-\d \t_3^{v} e^{i\w \t_3^{vs} \g_5} = \bigg\{ \begin{array}{ll}
		-\d e^{i\w \g_5}& \textrm{for the $up$ propagator} \\
		+\d e^{-i\w \g_5}& \textrm{for the $down$ propagator}
	\end{array}\, ,
\end{equation}
where $\w$ is determined from the given twisted-mass lattice action with $\d = 0$.  The second option is to add the mass term, assuming perfect maximal twist has been achieved;
\begin{equation}\label{eq:option2}
-i\d \mc{P}_V \g_5 = \bigg\{ \begin{array}{ll}
		-i\d \g_5 & \textrm{for the $up$ propagator} \\
		-i\d \g_5 & \textrm{for the $down$ propagator}
	\end{array} \, .
\end{equation}
While the second option may be simpler to implement numerically, the first option is slightly cleaner theoretically, although we emphasize again both are equivalent in the continuum limit.  We therefore proceed assuming option one~\eqref{eq:option1} has been implemented and will point out the differences if the second option were implemented.

Before proceeding, we examine the symmetry properties of this new mass operator.  The standard twisted mass action has an exact combined parity and flavor exchange symmetry, $\mc{P}_F^1$,
\begin{equation}
\mathcal{P}^1_F \colon
	\begin{cases}
	U_4(x) \rightarrow U_4(x_P) \,, \quad x_P = (-\mathbf{x},t) \\
	U_k(x) \rightarrow U^\dagger_k(x_P) \,, \quad k = 1,\,2,\,3 \\  
	\psi(x) \rightarrow i\tau_1 \gamma_4 \psi(x_P) \\
	\bar{\psi}(x) \rightarrow -i\bar{\psi}(x_P) \gamma_4 \tau_1
\end{cases} \, .
\end{equation}
The new mass term explicitly breaks this symmetry, 
\begin{equation}
	-i \d \bar{Q} \mc{P}_V \g_5 Q \longrightarrow +i \d \bar{Q} \mc{P}_V \g_5 Q\, .
\end{equation}
Thus, for example, the twisted mass term will now receive additive corrections.  However, this symmetry breaking is only present in the valence sector, and will not feed back into the gauge action.  Furthermore, this breaking must vanish as $\d \rightarrow 0$, and thus in non-parity violating matrix elements, the error must be of $\mc{O}(\d^2)$ or higher, including the additive correction to the twisted mass term, or the value of $\k_{critical}$.

\section{Partially quenched twisted mass chiral Lagrangian\label{sec:PQtmChPT}}
To understand the ramifications of this lattice action, we study the low energy theory with partially quenched $\chi$PT~\cite{Bernard:1992mk,Bernard:1993sv,Sharpe:1997by,Sharpe:2000bc,Sharpe:2001fh,Labrenz:1996jy,Chen:2001yi,Beane:2002vq}.  To begin, we must first construct the continuum Symanzik action at the quark level~\cite{Symanzik:1983dc,Symanzik:1983gh}, and then proceed to construct the resulting chiral Lagrangian~\cite{Sharpe:1998xm}.  Our proposed lattice action is given by%
\footnote{The ghost quarks are not actually computed in the numerical simulations, but are a theoretical construct to account for the lack of closed valence quark loops.} 
\begin{align}\label{eq:myAction}
\mc{S} = \sum_x & \bar{\psi}(x) \bigg\{ \frac{1}{2}\sum_\nu \left[ \g_\nu (\nabla_\nu^* + \nabla_\nu )
	-r\, \nabla_\nu^* \nabla_\nu \right]
	+ m_0 + i \g_5 \mu_0 \tau_3^{vs} 
	- \delta_0 \tau_3^{v} e^{i\w \t_3^{vs} \g_5} 
	\bigg\} \psi(x)
\end{align}
Expanding about the continuum limit, the effective continuum Lagrangian for the quarks can be written (to LO in the lattice spacing)
\begin{align}
\mc{L} = \bar{q} \Big[ \Dslash + \left(m - \d \t_3^v \right) e^{i\w \t_3^{vs} \g_5}  
	+a\, c_{sw}\, i \s_{\mu\nu} F_{\mu\nu} \Big] q\, . 
\end{align}
The renormalized quark mass is defined in the standard way with
\begin{align}
m^\prime &\equiv m \cos(\w) = Z_m (m_0 - m_c) / a\, ,&
\\
\mu &\equiv m \sin(\w) = Z_\mu \mu_0 / a\, .&
\end{align}
Here, $m_c$ is the critical mass (defined up to an $\mc{O}(a)$ shift).  The twisted mass term, $i\mu \t_3^{vs} \g_5$, is protected from additive mass corrections from the symmetries of the action (up to the $\mc{O}(\d^2)$ corrections mentioned above).  The twist angle is defined through the ratio $\tan(\w) = \mu / m^\prime$.  Similarly, the valence isospin breaking mass term will also be protected from additive mass renormalization (to all orders),
\begin{equation}
	\d = Z_\d \d_0 / a\, .
\end{equation}
There are additional operators at $\mc{O}(a)$, but these are suppressed by additional powers of the quark masses and will not modify the construction of the chiral Lagrangian through NLO~\cite{Sharpe:2004ny}.  The resulting partially quenched meson Lagrangian is given through $\mc{O}(m_q^2, m_q a, a^2)$ by
\begin{widetext}
\begin{align}\label{eq:PQtmChPT}
\mc{L}^{(PQ)} =&\ \frac{f^2}{8} \sTr \left( \partial_\mu \S \partial_\mu \S^\dagger \right)
	-\frac{f^2}{8} \sTr \left( \chi^{\prime\dagger} \S + \S^\dagger \chi^{\prime} \right)
	-L_1^{(PQ)} \left[ \sTr \left( \partial_\mu \S \partial_\mu \S^\dagger \right) \right]^2
	-L_2^{(PQ)} \sTr \left( \partial_\mu \S \partial_\nu \S^\dagger \right) \sTr \left( \partial_\mu \S \partial_\nu \S^\dagger \right)
\nonumber\\&
	-L_3^{(PQ)} \sTr \left( \partial_\mu \S \partial_\mu \S^\dagger \partial_\nu \S \partial_\nu \S^\dagger \right)
	+L_4^{(PQ)} \sTr \left( \partial_\mu \S \partial_\mu \S^\dagger \right) 
		\sTr \left( \chi^{\prime\dagger} \S + \S^\dagger \chi^{\prime} \right)
	-L_6^{(PQ)} \left[ \sTr \left( \chi^{\prime\dagger} \S + \S^\dagger \chi^{\prime} \right) \right]^2
\nonumber\\&
	+L_5^{(PQ)} \sTr \left( \partial_\mu \S \partial_\mu \S^\dagger 
		\left( \chi^{\prime\dagger} \S + \S^\dagger \chi^{\prime} \right) \right)
	-L_7^{(PQ)} \left[ \sTr \left( \chi^{\prime\dagger} \S - \S^\dagger \chi^{\prime} \right) \right]^2
	-L_8^{(PQ)} \sTr \left( \chi^{\prime\dagger} \S \chi^{\prime\dagger} \S 
		+ \S^\dagger \chi^{\prime} \S^\dagger \chi^{\prime}\right)
\nonumber\\&
	+W_4^{(PQ)} \sTr \left( \partial_\mu \S \partial_\mu \S^\dagger \right) 
		\sTr \left( \hat{A}^\dagger \S + \S^\dagger \hat{A} \right)
	+W_5^{(PQ)} \sTr \left( \partial_\mu \S \partial_\mu \S^\dagger  \left( \hat{A}^\dagger \S + \S^\dagger \hat{A} \right) \right)
\nonumber\\&
	-W_6^{(PQ)} \sTr \left( \chi^{\prime\dagger} \S + \S^\dagger \chi^{\prime} \right)
		\sTr \left( \hat{A}^\dagger \S + \S^\dagger \hat{A} \right)
	-W_7^{(PQ)}  \sTr \left( \chi^{\prime\dagger} \S - \S^\dagger \chi^{\prime} \right)
		\sTr \left( \hat{A}^\dagger \S - \S^\dagger \hat{A} \right)
\nonumber\\&
	-W_8^{(PQ)} \sTr \left( \chi^{\prime\dagger} \S \hat{A}^\dagger \S 
		+ \S^\dagger \chi^{\prime} \S^\dagger \hat{A} \right)
	-W_6^{\prime(PQ)} \left[ \sTr \left( \hat{A}^\dagger \S + \S^\dagger \hat{A} \right) \right]^2
	-W_8^{\prime(PQ)} \sTr \left( \hat{A}^\dagger \S \hat{A}^\dagger \S 
		+ \S^\dagger \hat{A} \S^\dagger \hat{A} \right)\, .
\end{align}
\end{widetext}
Here, we have already absorbed the leading discretization effects into a redefinition of $\chi$~\cite{Rupak:2002sm};
\begin{align}\label{eq:chiPrime}
	\chi^\prime &= 2B_0\left( m^\prime + i \mu \t_3^{vs} -\d \t_3^{v}e^{i\w \t_3^{vs}} \right) + 2W_0 a
\nonumber\\&
	\equiv \hat{m}^\prime + i \hat{\mu} \t_3^{vs} - \hat{\d} \t_3^{v}e^{i\w \t_3^{vs}} + \hat{a}
\nonumber\\
	\hat{A} &=2W_0 a \equiv \hat{a}
\end{align}
Further, we assume a power counting
\begin{equation}
\sqrt{(\hat{m}^\prime +\hat{a})^2 + \hat{\mu}^2}  \sim \hat{\d} \sim \hat{a}\, .
\end{equation}
The Lagrangian~\eqref{eq:PQtmChPT} is then the complete NLO meson Lagrangian relevant for our work.  To extract relevant physics information, this Lagrangian must be matched to its non partially quenched counterpart.  The unquenched Lagrangian is given by~\cite{Sharpe:2004ny}
\begin{widetext}
\begin{align}
\mc{L} =&\ \frac{f^2}{8} \Tr \left( \partial_\mu \S \partial_\mu \S^\dagger \right)
	-\frac{f^2}{8} \Tr \left( \chi^{\prime\dagger} \S + \S^\dagger \chi^{\prime} \right)
	-\frac{l_1}{4} \left[ \Tr \left( \partial_\mu \S \partial_\mu \S^\dagger \right) \right]^2
	-\frac{l_2}{4} \Tr \left( \partial_\mu \S \partial_\nu \S^\dagger \right) \Tr \left( \partial_\mu \S \partial_\nu \S^\dagger \right)
\nonumber\\&
	-\frac{l_3 + l_4}{16}  \left[ \Tr \left( \chi^{\prime\dagger} \S + \S^\dagger \chi^{\prime} \right) \right]^2
	+\frac{l_4}{8} \Tr \left( \partial_\mu \S \partial_\mu \S^\dagger \right) 
		\Tr \left( \chi^{\prime\dagger} \S + \S^\dagger \chi^{\prime} \right)
	-\frac{l_7}{16} \left[ \Tr \left( \chi^{\prime\dagger} \S - \S^\dagger \chi^{\prime} \right) \right]^2
\nonumber\\&
	+\tilde{W} \Tr \left( \partial_\mu \S \partial_\mu \S^\dagger \right) 
		\Tr \left( \hat{A}^\dagger \S + \S^\dagger \hat{A} \right)
	-W \Tr \left( \chi^{\prime\dagger} \S + \S^\dagger \chi^{\prime} \right)
		\Tr \left( \hat{A}^\dagger \S + \S^\dagger \hat{A} \right)
	-W^{\prime} \left[ \Tr \left( \hat{A}^\dagger \S + \S^\dagger \hat{A} \right) \right]^2
\nonumber\\&
	-W_7  \Tr \left( \chi^{\prime\dagger} \S - \S^\dagger \chi^{\prime} \right)
		\Tr \left( \hat{A}^\dagger \S - \S^\dagger \hat{A} \right)\, .
\end{align}
\end{widetext}
Here, the $l_i$ are the $SU(2)$ Gasser-Leutwyler coefficients~\cite{Gasser:1983yg}.  The matching of the continuum NLO Lagrangian has been performed~\cite{Bijnens:2005pa}, and here we also match the partially quenched twisted mass Lagrangian;
\begin{align}\label{eq:matching}
&l_1 = 4L_1^{(PQ)} +2L_3^{(PQ)}& 
&l_2 = 4L_2^{(PQ)}& \nonumber\\
&l_3 + l_4 = 16L_6^{(PQ)} +8L_8^{(PQ)}& 
&l_4 = 8L_4^{(PQ)} +4L_5^{(PQ)}& \nonumber\\
&l_7 = 16L_7^{(PQ)} +8L_8^{(PQ)}& \nonumber\\
&\tilde{W} = W_4^{(PQ)} +\frac{1}{2}W_5^{(PQ)}& 
&W = W_6^{(PQ)} +\frac{1}{2}W_8^{(PQ)}& \nonumber\\
&W_7 = W_7^{(PQ)} +\frac{1}{2}W_8^{(PQ)}&
&W^\prime = W_6^{\prime(PQ)} +\frac{1}{2}W_8^{\prime(PQ)}&
\end{align}

\subsection{Vacuum Alignment}
To determine properties of the low energy theory, it is useful to first expand about the vacuum of the theory.  In unquenched mass-degenerate twisted mass $\chi$PT, it has been shown that the vacuum is given at NLO by~\cite{Sharpe:2004ny}
\begin{equation}
\S_0 \equiv \langle 0 | \S | 0 \rangle
	= \textrm{exp}(i\w \t_3)\, ,
\end{equation}
where $\w = \w_0 + \e$ and
\begin{align}\label{eq:tmvac}
\textrm{exp}(i\w_0 \t_3) &= \frac{\hat{m}^\prime + \hat{a} + i\hat{\mu}\t_3}{\hat{m}}\, ,
\nonumber\\
\e(\w) &= -\frac{32}{f^2}\hat{a} \sin(\w) 
	\left( W + 2W^\prime \cos(\w) \frac{\hat{a}}{\hat{m}} \right)\, ,
\nonumber\\
\hat{m} &= \sqrt{(\hat{m}^\prime + \hat{a})^2 + \hat{\mu}^2}\, .
\end{align}
In the continuum limit, $\e \rightarrow 0$ as one would expect.

In our proposed partially quenched theory, the vacuum has a slight additional perturbation.  To find the vacuum, one can either minimize the potential energy (which is slightly more complicated for a partially quenched theories~\cite{Golterman:2005ie}) or one can require all single-pion vertices to vanish.  Working through NLO, one can show that the Lagrangian, as written in Eq.~\eqref{eq:PQtmChPT}, is rotated from the vacuum by the following angle
\begin{align}
	\S &= \xi_m \S_{ph} \xi_m\quad \textrm{with} \nonumber\\
	\xi_m &= \textrm{exp} \left[ \frac{i}{2} \left(\w \t_3^{vs} + \e^\prime \t_3^{v} + \e^{\prime\prime}\mc{P}_V \right) \right]
\end{align}
where 
\begin{align}\label{eq:tmPQvac}
&\w = \w_0 + \e(\w)\, ,& \nonumber\\
&\e^\prime(\w) = \frac{\hat{\d}^2}{\hat{m}^2 - \hat{\d}^2}
	\left[ \e(\w) + \frac{16 \hat{a}\sin(\w)W_8^{(PQ)}}{f^2} \right] \, ,&\nonumber\\
&\e^{\prime\prime}(\w) = \frac{\hat{\d}\hat{m}}{\hat{m}^2 - \hat{\d}^2}
	\left[ \e(\w) + \frac{16 \hat{a}\sin(\w)W_8^{(PQ)}}{f^2} \right] \, ,
\end{align}
and $\hat{m}$ and $\e(\w)$ are given by Eq.~\eqref{eq:tmvac}, and 
\begin{align}
&\S_{ph} = \textrm{exp} \left( \frac{2i \Phi}{f} \right), \quad \textrm{with}&
\nonumber\\
&\Phi=\begin{pmatrix}
			\eta_u & \pi^+ & \phi_{uj} & \phi_{ul} & \phi_{u\tilde{u}}& \phi_{u\tilde{d}} \\
			\pi^- & \eta_d & \phi_{dj} & \phi_{dl} & \phi_{d\tilde{u}} &\phi_{d\tilde{d}} \\
			\phi_{ju} &\phi_{jd} & \eta_j & \phi_{jl} & \phi_{j\tilde{u}} & \phi_{j\tilde{d}} \\
			\phi_{lu} &\phi_{ld} & \phi_{lj} & \eta_l  & \phi_{l\tilde{u}} & \phi_{l\tilde{d}} \\
			\phi_{\tilde{u}u} &\phi_{\tilde{u}d} & \phi_{\tilde{u}j} & \phi_{\tilde{u}l} & \tilde{\eta}_u & \phi_{\tilde{u}\tilde{d}} \\
			\phi_{\tilde{d}u} &\phi_{\tilde{d}d} & \phi_{\tilde{d}j} & \phi_{\tilde{d}l} & \phi_{\tilde{d}\tilde{d}} &\tilde{\eta}_d \\
		\end{pmatrix}\, .&
\end{align}
Again, in the continuum limit, one finds the extra rotations of the vacuum vanish as expected.  If we had used \textit{option 2}, Eq.~\eqref{eq:option2}, then both $\e^\prime$ and $\e^{\prime\prime}$ would have an additional $\mc{O}(a)$ shift from the tuning of $\w$.

\subsection{Partially Quenched HairPin Interactions}

In the partially quenched Lagrangian, there are additional operators we have not written in Eq.~\eqref{eq:PQtmChPT}, related to the singlet field $\Phi_0 = \str(\Phi) / \sqrt{N_F}$, which will ultimately be integrated out of the theory.  These operators can be included to determine the structure of the hairpin interactions which give rise to the partially-quenched chiral logarithms~\cite{Sharpe:1997by,Sharpe:2000bc,Sharpe:2001fh}.  When calculating properties of pions, it is convenient to use the basis of fields~\cite{Chen:2005ab,Chen:2006wf},%
\footnote{Keeping the mass of the singlet explicit ($m_0$), these two fields are given at leading order in $\hat{\d}$ by $\pi^0 = \frac{\eta_u - \eta_d}{\sqrt{2}} + \frac{\hat{\d}}{2m_0^2}\frac{\eta_u + \eta_d}{\sqrt{2}}$ and  $\bar{\eta} = \frac{\eta_u + \eta_d}{\sqrt{2}} - \frac{\hat{\d}}{2m_0^2}\frac{\eta_u - \eta_d}{\sqrt{2}}$ as the isospin breaking mass term allows a coupling between the $\pi^0$ and $\eta^\prime$ fields.  These extra terms need not be included in the propagators as they are included as interactions in the partially quenched chiral Lagrangian.}
\begin{align}
	&\pi^0 = \frac{\eta_u - \eta_d}{\sqrt{2}}\, ,& 
	&\bar{\eta} = \frac{\eta_u + \eta_d}{\sqrt{2}}\, .&
\end{align}
One can show the propagators of these fields, which can be determined from the Appendix of Ref.~\cite{Chen:2007ug} (this work considered a mixed action with twisted mass one of the possible types of sea quarks),
\begin{align}\label{eq:etaProps}
	&\mc{G}_{\pi^0} = \frac{1}{p^2 + \hat{m}}\, ,&
	&\mc{G}_{\bar{\eta}} = \frac{-\hat{\d}^2}{(p^2 + \hat{m})^3}\, .&
\end{align}
With these propagators, one can determine the loop corrections to the pion masses and decay constants.  The resulting loop corrections are given by
\begin{equation}
(\d m_\pi^2)^{loop} = \frac{\hat{m}}{f^2} i \mc{I}(\hat{m}) 
	+\frac{\hat{m}\hat{\d}^2}{2 f^2} \left( \frac{\partial}{\partial \hat{m}} \right)^2 i\mc{I}(\hat{m})\, ,
\end{equation}
where the loop function is defined as the regulated 4d Euclidean integral
\begin{equation}
	i\mc{I}(\hat{m}) \equiv \int_R \frac{d^4 q_E}{(2\pi)^4} \frac{1}{q_E^2 + \hat{m}}\, ,
\end{equation}
which in the \textit{modified dimensional regularization} of Ref.~\cite{Gasser:1983yg} is
\begin{equation}
	i\bar{\mc{I}}(\hat{m}) = \frac{\hat{m}}{(4\pi)^2} \ln \left( \frac{\hat{m}}{\L^2} \right)\, .
\end{equation}
This expression, with the known volume corrections to the one-loop pion mass formula~\cite{Gasser:1986vb,Colangelo:2003hf}
\begin{equation}
i\mc{I}[FV] - i\mc{I}[\infty V] = \frac{m_\pi^2}{4\pi^2} \frac{1}{m_\pi L} \sum_{\vec{n} \neq 0} \frac{1}{|\vec{n}|} K_1(|\vec{n}| m_\pi L)\, ,
\end{equation}
can be used to trivially determine the volume corrections from the hairpin contribution (we have made use of the consistency at NLO of replacing the leading order pion mass with the full pion mass in these expressions, $\hat{m} \rightarrow m_\pi^2$).

\subsection{Pion Masses\label{sec:PionMasses}}
Using this Lagrangian, we can now determine the pion masses.  Working through $\mc{O}(m_q^2,m_qa,a^2)$, and using the matching relations Eq.~\eqref{eq:matching}, we find
\begin{align}
	m_{\pi^\pm}^2 =&\ \hat{m} \left[ 1 + \frac{\hat{m}}{(4\pi f)^2} \ln \left( \frac{\hat{m}}{\L^2} \right) 
		+ \frac{4\hat{m}}{f^2} l_3^r(\L) \right]
		+\frac{\D_{PQ}^4}{2(4\pi f)^2}
		+\frac{32 \hat{a}\cos(\w)}{f^2} \left[ \hat{m}(2W -\tilde{W}) +2W^\prime \hat{a}\cos(\w) \right]\, ,
\nonumber\\
	m_{\pi^0}^2 =&\  m_{\pi^\pm}^2 + \frac{4\hat{\d}^2}{f^2} l_7 -\frac{64W^\prime \hat{a}^2 \sin^2(\w)}{f^2}\, .
\end{align}
The first non-standard term in this expression arises from the hairpin interactions, and is the remnant of the partially quenched chiral Log.  Because the $\bar{\eta}$ propagator has extra suppression, Eq.~\eqref{eq:etaProps} relative to the standard version of this partially quenched propagator~\cite{Chen:2005ab}, (this happens because $\hat{m}_{val} = \hat{m}_{sea}$), the enhanced chiral logarithm has become simply a constant.  To clarify which contributions are physical, and which are partially quenched artifacts, we have introduced the term
\begin{equation}\label{eq:DPQ}
	\D_{PQ}^2 = \hat{\d}\, ,
\end{equation}
which we shall use in the remaining mass expressions.  For $\hat{\d} \rightarrow 0$ these expressions reduce to those of standard twisted mass $\chi$PT.  Therefore, with multiple values of the isospin breaking mass term $\d$, one can determine $l_7$ from the charged-neutral pion mass splitting.
At this order, the pion decay constants do not receive any corrections and are given by the standard form
\begin{align}
f_{\pi} =&\ f \bigg[ 
	1 - \frac{2\hat{m}}{(4\pi f)^2} \ln \left(\frac{\hat{m}}{\L^2} \right)
	+\frac{2 \hat{m}}{f^2}l_4^r(\L) 
	+\cos\w \frac{16\tilde{W}\hat{a}}{f^2} \bigg]\, .
\end{align}

\subsection{Baryons\label{sec:baryonMasses}}
One can also include baryons in twisted mass $\chi$PT~\cite{WalkerLoud:2005bt}, using an extension of the heavy baryon chiral Lagrangian formulated by Jenkins and Manohar~\cite{Jenkins:1990jv,Jenkins:1991ts}.  For our work, we will need the two-flavor partially quenched baryon Lagrangian, which was first developed in Ref.~\cite{Beane:2002vq} and later extended to NNLO in Ref.~\cite{Tiburzi:2005na}.  Here we use the normalization conventions of Ref.~\cite{Tiburzi:2008bk}, for which the twisted mass baryon chiral Lagrangian is given by
\begin{widetext}
\begin{align} \label{eq:LOPQBaryons}
\mc{L}^{(PQ)} =&\ 
	\left( \ol{\mc{B}} v \cdot D \mc{B} \right) 
	+\frac{\a_M^{(PQ)}}{(4\pi f)}\,\left( \ol{\mc{B}} \mc{B} \mc{M}_+ \right)
	+\frac{\b_M^{(PQ)}}{(4\pi f)}\,\left( \ol{\mc{B}} \mc{M}_+ \mc{B}\right)
	+\frac{\s_M^{(PQ)}}{(4\pi f)}\,\left( \ol{\mc{B}} \mc{B} \right) \,{\rm tr}(\mc{M}_+)
	+\frac{\s_W^{(PQ)}}{(4\pi f)}\,\left( \ol{\mc{B}} \mc{B} \right)\,{\rm tr}(\mc{W}_+) \notag \\
	& -(\ol{\mc{T}}_\mu v \cdot D\,\mc{T}_\mu) 
	- \Delta\,(\ol{\mc{T}}_\mu \mc{T}_\mu) 
	+\frac{\g_M^{(PQ)}}{(4\pi f)}\,(\ol{\mc{T}}_\mu \mathcal{M}_+ \mc{T}_\mu)
	-\frac{\ol{\s}_M^{(PQ)}}{(4\pi f)}\,(\ol{\mc{T}}_\mu \mc{T}_\mu)\,{\rm tr}(\mathcal{M}_+)  
	-\frac{\ol{\s}_W^{(PQ)}}{(4\pi f)}\,(\ol{\mc{T}}_\mu \mc{T}_\mu)\,{\rm tr}(\mathcal{W}_+)
\nonumber\\&\ 
	+2\a^{(PQ)}\left(\ol{\mc{B}} S^\mu \mc{B} A_\mu \right)\ 
	+\ 2\b^{(PQ)} \left(\ol{\mc{B}} S^\mu A_\mu \mc{B} \right)\ 
	+\ 2{\mc{H}^{(PQ)}} \left( \ol{\mc{T}} {}^\nu S^\mu A_\mu \mc{T}_\nu \right)
     +\sqrt{\frac{3}{2}}{\mc{C}} \left[ \left( \ol{\mc{T}}{}^\nu A_\nu \mc{B}
      \right)\ 
    +\ \left( \ol{\mc{B}} A_\nu \mc{T}^\nu \right) \right]. 
\end{align}
\end{widetext}
In this Lagrangian, $(\ )$ denote the graded summation of flavor indices, first defined in Ref.~\cite{Labrenz:1996jy}.  The spurions are defined as
\begin{align}
\mc{M}_\pm &= \frac{1}{4} \left( \xi \chi^{\prime^\dagger} \xi \pm \xi^\dagger \chi^\prime \xi^\dagger \right)\, ,
\nonumber\\
\mc{W}_\pm &= \frac{1}{4} \left( \xi \hat{A}^\dagger \xi \pm \xi^\dagger \hat{A} \xi^\dagger \right)\, ,
\end{align}
with $\chi^\prime$ and $\hat{A}$ defined in Eq.~\eqref{eq:chiPrime}.  Here, $\xi^2 = \S$ is needed for the inclusion of the baryon fields in the chiral Lagrangian.  The axial field is defined as
\begin{equation}
A_\mu = \frac{i}{2} \left( \xi \partial_\mu \xi^\dagger - \xi^\dagger \partial_\mu \xi \right)\, ,
\end{equation}
and $S_\mu$ is a spin operator~\cite{Jenkins:1990jv,Jenkins:1991ts}.  As with the mesons, we must match this Lagrangian to the unquenched one, given by
\begin{widetext}
\begin{align} \label{eq:LOBaryons}
\mc{L} =&\ 
	\ol{N} v \cdot D N 
	+\frac{\a_M}{(4\pi f)}\,\ol{N} \mc{M}_+ N
	+\frac{\s_M}{(4\pi f)}\,\ol{N} N \,{\rm tr}(\mc{M}_+)
	+\frac{\s_W}{(4\pi f)}\,\ol{N} N \,{\rm tr}(\mc{W}_+) 
\nonumber\\&\ 
	+ (\ol{T}_\mu v \cdot D\,T_\mu) 
	+ \Delta\,(\overline{T}_\mu T_\mu) 
	+\frac{\g_M}{(4\pi f)}\,(\overline{T}_\mu \mathcal{M}_+ T_\mu)
	+\frac{\ol{\s}_M}{(4\pi f)}\,(\ol{T}_\mu T_\mu)\,{\rm tr}(\mathcal{M}_+)  
	+\frac{\ol{\s}_W}{(4\pi f)}\,(\ol{T}_\mu T_\mu)\,{\rm tr}(\mathcal{W}_+)
\nonumber\\&\ 
	+2\,g_A \,\overline{N} S \cdot {A} \, N 
	-2\,g_{\Delta\Delta}\,\ol{T}_\mu S\cdot {A}\,T_\mu  
	+g_{\Delta N}\,\left[\ol{T}^{kji}_\mu {A}_i^{\mu,i^\prime}
                    \epsilon_{j i^\prime} N_k + h.c.\right] \, .
\end{align}
\end{widetext}
Performing the matching, one finds
\begin{align}\label{eq:baryonMatch}
&\a_M = \frac{2}{3}\a_M^{(PQ)} - \frac{1}{3}\b_M^{(PQ)},&
\nonumber\\
&\s_M = \s_M^{(PQ)} +\frac{1}{6}\a_M^{(PQ)} + \frac{2}{3}\b_M^{(PQ)},&
\nonumber\\
&\g_M = \g_M^{(PQ)}\ ,\ \bar{\s}_M = \bar{\s}_M^{(PQ)},&
\nonumber\\
&g_A = \frac{2}{3}\a^{(PQ)} - \frac{1}{3}\b^{(PQ)}\ ,\ 
	g_1 = \frac{1}{3}\a^{(PQ)} + \frac{4}{3}\b^{PQ)},&
\nonumber\\
&g_{\D\D} = \mc{H}\ ,\ 
	g_{\D N} = -\mc{C},&
\end{align}
and
\begin{equation}\label{eq:baryonTMmatch}
\s_W = \s_W^{(PQ)}\ ,\ 
	\bar{\s}_W = \bar{\s}_W^{(PQ)}\, ,
\end{equation}
for the discretization operators.  Using the partially quenched Lagrangian, performing the matching with Eqs.~\eqref{eq:baryonMatch} and \eqref{eq:baryonTMmatch}, and working consistently to NLO we find the nucleon masses are given by
\begin{widetext}
\begin{align}
m_p =&\ M_0 
	-\frac{\hat{\d}}{(4\pi f_\pi)} \frac{\a_M}{2}
	+ \frac{m_\pi^2}{(4\pi f_\pi)} \left(\frac{\a_M}{2} + \s_M^r(\L) \right)
	-\frac{3\pi g_A^2}{(4\pi f_\pi)^2}\, m_\pi^3
	-\frac{8g_{\D N}^2}{3(4\pi f_\pi)^2}\mc{F}(m_\pi,\D,\L)
\nonumber\\&\ 
	+\frac{\hat{a}\cos(\w)\s_W}{(4\pi f_\pi)}
	+\frac{3\pi \D_{PQ}^4(g_A + g_1)^2}{8m_\pi (4\pi f_\pi)^2}
\nonumber\\
m_n =&\ M_0 
	+\frac{\hat{\d}}{(4\pi f_\pi)} \frac{\a_M}{2}
	+ \frac{m_\pi^2}{(4\pi f_\pi)} \left(\frac{\a_M}{2} + \s_M^r(\L) \right)
	-\frac{3\pi g_A^2}{(4\pi f_\pi)^2}\, m_\pi^3
	-\frac{8g_{\D N}^2}{3(4\pi f_\pi)^2}\mc{F}(m_\pi,\D,\L)
\nonumber\\&\ 
	+\frac{\hat{a}\cos(\w)\s_W}{(4\pi f_\pi)}
	+\frac{3\pi \D_{PQ}^4(g_A + g_1)^2}{8m_\pi (4\pi f_\pi)^2}
\nonumber\\
&\ \textrm{with}
\nonumber\\
\mc{F}(m,\D,\L) =&\
	(\D^2 - m^2 +i\e)^{3/2}
	\ln \left( \frac{\D+\sqrt{\D^2 - m^2 + i\e}}{\D - \sqrt{\D^2 - m^2 + i\e}} \right)
	-\frac{3}{2}\D m^2 \ln \left( \frac{m^2}{\L^2}\right)
	-\D^3 \ln \left(\frac{4\D^2}{m^2}\right)\, .
\end{align}
\end{widetext}
Here, we see that the NLO contributions exactly cancel in $m_n - m_p$, thus rendering the expansion of the mass splitting on the same footing as the pion mass expansion.  One can see the last term in these mass expressions is proportional to the coupling of the nucleons to the singlet field, being proportional to $(g_A+g_1)$.  These terms are remnants of our partially quenched theory and would vanish if the sea quarks had an isospin breaking mass term equal to that of the valence quarks.

Similarly, one can determine the delta mass expressions.  One should caution that due to the strong coupling to the $\pi-N$ system, the deltas, at lighter pion masses, have significantly larger volume dependence than the nucleons or pions~\cite{Luscher:1991cf,Bernard:2008zza,Bernard:2008ax}.  Neglecting these issues for this work, the delta mass extrapolation formulae are given by
\begin{widetext}
\begin{align}
m_\D =&\ M_0 + \D 
	+\frac{\hat{\d}}{(4\pi f_\pi)}\frac{c_{\D}\, \g_M}{6} 
	+\frac{m_\pi^2}{(4\pi f_\pi)} \left( \frac{\g_M}{2} + \bar{\s}_M^r(\L) \right)
	-\frac{25 g_{\D\D}^2}{27(4 \pi f_\pi)^2} \, m_\pi^3 
	- \frac{2 g_{\D N}^2}{3(4 \pi f)^2} \, \mc{F} (m_\pi,-\D,\mu)
\nonumber\\&\ 
	+\frac{\hat{a}\cos(\w)\bar{\s}_W}{(4\pi f_\pi)}
	+\frac{5\pi \D_{PQ}^4 g_{\D\D}^2}{12m_\pi(4\pi f_\pi)^2}\, ,
\end{align}
\end{widetext}
where the coefficients $c_{\D}$ are given in Table~\ref{tab:cD} and 
\begin{align}
\mc{F}(m,-\D,\L)&= \left\{ \begin{array}{lc}
	-\mc{F}(m,\D,\L) -2i\pi (\D^2 - m^2)^{3/2}, & m < |\D| \\
	-\mc{F}(m,\D,\L) -2\pi (m^2 - \D^2)^{3/2}, & m > |\D|
	\end{array} \right. \, .
\end{align}
In the limit $\hat{\d} \rightarrow 0$, these expressions reduce to those in Ref.~\cite{WalkerLoud:2005bt}.  Similar to the nucleons, the NLO contributions exactly cancel from the mass splittings, and the last term in this expression arises from the partially quenched construction.  Also, at this order, one sees the delta masses obey an equal spacing rule, which is violated at NNLO by one operator~\cite{Tiburzi:2005na}.  It is precisely the imaginary piece of this $\mc{F}$-function, which in finite Euclidean volume gives rise to the power-law dependence of the delta masses~\cite{Bernard:2008zza,Bernard:2008ax}.

\begin{table}[t]
\caption{\label{tab:cD}\textit{Coefficients of the delta mass corrections arising from the LO  term proportional to the strong isospin breaking and the NNLO discretization effect.}}
\begin{ruledtabular}
\begin{tabular}{ccccc}
& $\D^{++}$ & $\D^{+}$& $\D^{0}$& $\D^{-}$\\
\hline
$c_{\D}$ & -3& -1& 1& 3 \\
$\tilde{c}_{\D}$ & 1& -1/3& -1/3& 1
\end{tabular}
\end{ruledtabular}
\end{table}

\subsubsection{NNLO}
For the baryons, we can carry this prescription to NNLO (which is $\mc{O}(m_q^2, m_q a, a^2))$.  The complete set of partially quenched operators relevant for the nucleon and delta masses at $\mc{O}(m_q^2)$ was determined in Ref.~\cite{Tiburzi:2005na}.  The complete set of new twisted mass heavy baryon $\chi$PT operators at $\mc{O}(m_q a, a^2)$ was determined in Ref.~\cite{WalkerLoud:2005bt}.  The extension of the twisted mass operators to the partially quenched theory is straightforward, however due to the cumbersome length, we do not detail them here.  Rather we present the results of the mass corrections after the matching to the unquenched theory has been performed, and we only provide the expressions at maximal twist.  For the nucleons, there are 9 relevant operators in the partially quenched theory, which reduce to four in the unquenched theory, while for the deltas there are 7 relevant operators in both,
\begin{align}
\mc{L}_M = \frac{1}{(4\pi f)^3} \bigg\{&
	b_1^M \bar{N}  \, \mc{M}_+^2  N
		+ b_5^M \bar{N} N \, \tr ( \mc{M}_+^2 )
		+ b_6^M \bar{N} \, \mc{M}_+ N \, \tr (\mc{M}_+) 
		+ b_8^M \bar{N} N \, [\tr (\mc{M}_+)]^2
\nonumber\\&
	+ t_1^M \, \bar{T} {}^{kji}_\mu (\mc{M}_+ \mc{M}_+)_{i}{}^{i'} T_{\mu, i'jk} 
	+ t_2^M  \, \bar{T} {}^{kji}_\mu (\mc{M}_+)_{i}{}^{i'} (\mc{M}_+)_{j}{}^{j'} T_{\mu, i'j'k} 
	+ t_3^M  \, \bar{T}_\mu T_\mu \tr (\mc{M}_+^2)
\nonumber\\&			
	+ t_4^M \, \left( \bar{T}_\mu \mc{M}_+ T_\mu \right) \tr (\mc{M}_+)
	+ t_5^M  \, \bar{T}_\mu T_\mu  [ \tr(\mc{M}_+) ]^2
\nonumber\\&
	+b_1^{W_-} \bar{N} N \Tr (W_- W_-)
	+t_1^{W_-} (\bar{T}_\mu T_\mu) \Tr (W_- W_-)
	+t_2^{W_-} \bar{T}^{kji}_\mu (W_-)_{i}^{\ i^\prime} (W_-)_{j}^{\ j^\prime} T_{\mu, i^\prime j^\prime k}
	\bigg\}\, .
\end{align}
As discussed in Refs.~\cite{WalkerLoud:2005bt,AbdelRehim:2005gz}, the symmetries of the twisted mass lattice action prevent the twisted mass term from splitting the nucleon masses, which is reflected in the low energy chiral Lagrangian, while the delta-multiplet splits into two degenerate pairs~\cite{WalkerLoud:2005bt}.  
The corrections to the delta masses from the twisted mass discretization, at maximal twist, are given by
\begin{equation}
 \d M_\D = -\frac{\hat{a}^2t_2^{W_-}}{4(4\pi f_\pi)^3} \tilde{c}_\D\, ,
 \end{equation}
where $\tilde{c}_\D$ is given in Table~\ref{tab:cD}.

With our particular partially quenched construction, the nucleon and delta masses will have an error at this order from one operator each, proportional to the terms with $\Tr (\mc{M}_+^2)$.  In the full theory with isospin breaking in both the sea and valence sector, the masses would receive mass corrections
\begin{align}
	\d m_N &= \frac{b_5^M(m_\pi^4 + \hat{\d}^2)}{2(4\pi f_\pi)^3}\, ,
\nonumber\\
	\d m_\D &= \frac{t_3^M(m_\pi^4 + \hat{\d}^2)}{2(4\pi f_\pi)^3}\, ,
\end{align}
while in our partially quenched theory, these mass corrections become
\begin{align}
	\d m_N &\rightarrow \frac{b_5^M(m_\pi^4)}{2(4\pi f_\pi)^3}\, ,
\nonumber\\
	\d m_\D &\rightarrow \frac{t_3^M(m_\pi^4)}{2(4\pi f_\pi)^3}\, .
\end{align}
However, in the mass splittings, these contributions vanish leaving the baryon mass splittings free of strong isospin breaking errors at this order.  At maximal twist, we then find the nucleon mass splittings is given through NNLO by the expression
\begin{align}\label{eq:MNsplit}
m_n - m_p =&\ \frac{\hat{\d}}{(4\pi f_\pi)} \bigg\{ \a_M
	+\frac{m_\pi^2}{(4\pi f_\pi)^2} (b_1^M + b_6^M)
	+\frac{\mc{J}(m_\pi,\D,\L)}{(4\pi f_\pi)^2} 4g_{\D N}^2 \left( \frac{5}{9}\g_M - \a_M \right)
\nonumber\\&\ 
	+ \frac{m_\pi^2}{(4\pi f_\pi)^2}\bigg[ \frac{20}{9}\g_M g_{\D N}^2 - 4\a_M (g_A^2 +g_{\D N}^2)
	-\a_M(6g_A^2 + 1) \ln \left( \frac{m_\pi^2}{\L^2} \right) \bigg]
\nonumber\\&\ 
	+\frac{\a_M\D_{PQ}^4}{m_\pi^2(4\pi f_\pi)^2} \left( 2 - \frac{3}{2}(g_A + g_1)^2 \right)
	\bigg\}\, ,
\end{align}
where
\begin{equation}
\mc{J}(m,\D,\L) = 
	2\D\sqrt{\D^2 - m^2+i\e} \ln \left( \frac{\D+\sqrt{\D^2 - m^2 + i\e}}{\D - \sqrt{\D^2 - m^2 + i\e}} \right)
	+m^2 \ln \left( \frac{m^2}{\L^2} \right)
	+2\D^2 \ln \left( \frac{4\D^2}{m^2} \right)\, .
\end{equation}
Note, the nucleon mass splitting is free of discretization errors at this order.  The last term in this mass splitting expression is not physical, having arisen from our partially quenched construction with $\D_{PQ}^2 = \hat{\d}$, Eq.~\eqref{eq:DPQ}.

Of phenomenological interest, the delta mass operator with coefficient $t_2^M$ leads to a breaking of the delta equal mass spacing~\cite{Tiburzi:2005na}.  Taking the delta mass splittings between neighboring members of the multiplet, one finds (at maximal twist)
\begin{align}\label{eq:MDsplit}
m_\D - m_{\D^\prime} =&\ \frac{\hat{\d}}{(4\pi f_\pi)} \bigg\{ \frac{\g_M}{3}
	+\frac{m_\pi^2}{(4\pi f_\pi)^2} \frac{t_1^M + t_2^M + t_3^M}{3}
	-\frac{\mc{J}(m_\pi, -\D,\L)}{(4\pi f_\pi)^2} \frac{g_{\D N}^2(2\g_M -\a_M)}{3}
\nonumber\\&\ 
	+\frac{m_\pi^2}{(4\pi f_\pi)^2} \left[ 
		\frac{c_{t_2}t_2^M}{3}
		-\frac{52}{81} \frac{\g_M}{3} g_{\D\D}^2
		-\frac{\g_M}{3}\left(1 + \frac{10g_{\D\D}^2}{27} \right) \ln \left( \frac{m_\pi^2}{\L^2} \right)
	\right]
\nonumber\\&\ 
	+\frac{\g_M \D_{PQ}^4}{18m_\pi^2(4\pi f_\pi)^2}\left( 12 - 5g_{\D\D}^2 \right)
	\bigg\}
	+\frac{c_{t_2} t_2^{W_-}}{3} \frac{\hat{a}^2}{(4\pi f_\pi)^3}
\end{align}
where
\begin{align}
\mc{J}(m,-\D,\L) &= \left\{ \begin{array}{lc}
	\mc{J}(m,\D,\L) +4i\pi \D \sqrt{\D^2-m^2},& m < |\D| \\
	\mc{J}(m,\D,\L) -4\pi \D \sqrt{m^2 - \D^2}, & m > |\D|
	\end{array}\right. \, ,
\end{align}
and $c_{t_2}$ is given in Table~\ref{tab:DmnsDprime}.
In the limit $\hat{\d} \rightarrow 0$, the delta mass splitting reduces to that in Ref.~\cite{WalkerLoud:2005bt},
\begin{equation}
	m_\D - m_{\D^\prime} = \frac{c_{t_2} t_2^{W_-}}{3} \frac{\hat{a}^2}{(4\pi f_\pi)^3}\, .
\end{equation}

\begin{table}[t]
\caption{\label{tab:DmnsDprime}\textit{Coefficients of the delta equal mass splitting violation.}}
\begin{ruledtabular}
\begin{tabular}{cccc}
& $m_{\D^+} - m_{\D^{++}}$ & $m_{\D^0} - m_{\D^{+}}$& $m_{\D^-} - m_{\D^{0}}$ \\
\hline
$c_{t_2}$ & -1& 0& 1 \\
\end{tabular}
\end{ruledtabular}
\end{table}

\section{Discussion\label{sec:Discussion}}
In this work, we have proposed a partially quenched extension of the twisted mass lattice action which includes a strong isospin breaking mass term in the valence sector, and preserves flavor identification.  Further, we developed the corresponding partially quenched twisted mass chiral perturbation theory relevant for our proposed lattice action.  While this proposal does not completely incorporate the effects of the strong isospin breaking, we have demonstrated, using the partially quenched theory, that the error in this approximation is sub-leading.  
This method extends to the hyperon spectrum in a straightforward manner.  The relevant partially quenched twisted mass $\chi$PT could be created by adding the isospin breaking mass term to the valence quarks in Refs.~\cite{Chen:2001yi,WalkerLoud:2004hf,Tiburzi:2004rh}, and then matched to the unquenched theory~\cite{Frink:2004ic}.  Alternatively, one could develop the partially quenched version of the two-flavor $\chi$PT for hyperons~\cite{Tiburzi:2008bk}.
One application of this work would be the calculation of a isospin violations in the hadron spectrum utilizing twisted mass LQCD, for which these errors are even further suppressed.  Combined with the physical point, this would allow for a lattice determination of the physical light quark mass splitting, $2\d = m_d - m_u$, a stated challenge in Ref.~\cite{Wilczek:2002wi}.

\bigskip
To contribute to this proposed endeavor, we have determined the relevant chiral extrapolation formula for the mass splittings of the pions, nucleons and deltas (as well as expressions for the masses themselves).  The leading non-vanishing correction to the pion mass splitting is
\begin{equation}
	m_{\pi^0}^2 - m_{\pi^\pm}^2 = \frac{4\hat{\d}^2}{f_\pi^2} l_7 - \frac{64W^\prime \hat{a}^2 \sin^2(\w)}{f_\pi^2}\, .
\end{equation}
At a fixed lattice spacing, with multiple values of the isospin breaking mass parameter, one can determine the Gasser-Leutwyler coefficient, $l_7$.  Using the physical values of the charged and neutral pion masses, after correcting for electromagnetic interactions, this would allow for a lattice determination of the $up$ and $down$ quark masses.  We have also provided expressions for the nucleon and delta mass splittings, and shown that through $\mc{O}(\hat{m}^2,\hat{\d}^2)$, these mass splittings are free from errors arising from the degeneracy of the sea quarks.  Scaling the masses by $f_\pi$, one can perform an extrapolation of these mass splittings which requires no scale setting.  The leading-order splittings (the NLO corrections exactly cancel and the full NNLO expressions are in Eqs.~\eqref{eq:MNsplit} and \eqref{eq:MDsplit} respectively),
\begin{align}
	\frac{m_n - m_p}{f_\pi} &= \frac{\a_M}{4\pi} \frac{\hat{\d}}{f_\pi^2} +\mc{O}(\hat{\d}m_\pi^2)\, ,
\nonumber\\
	\frac{m_\D - m_{\D^\prime}}{f_\pi} &= \frac{\g_M}{12\pi} \frac{\hat{\d}}{f_\pi^2} +\mc{O}(\hat{\d}m_\pi^2)\, .
\end{align}
After correcting for electromagnetic effects, these baryon mass splittings, combined with the physical mass splittings, could also be used to determine the physical mass splitting of the light quarks (perhaps we should caution that for present lattice simulations, the nucleon mass behaves linearly in $m_\pi$~\cite{WalkerLoud:2008bp,WalkerLoud:2008pj}, a trend which is not presently understood theoretically, and may complicate the baryon extrapolation).  Either way, one must determine the value of $\hat{\d}$, both for a given set of lattice input parameters and at the physical point.  A direct lattice calculation of this quantity requires the determination of the renormalization coefficient, $Z_\d$.  However, one can make a good approximation of this parameter by taking advantage of the fact both the twisted mass parameter as well as the isospin breaking mass parameter are protected from additive mass renormalization (actually, the explicit breaking of the parity-flavor symmetry by the proposed isospin breaking mass parameter gives rise to an $\mc{O}(\d^2)$ correction to $\mu$, but this is expected to be perturbative). From Eq.~\eqref{eq:chiPrime}, one has
\begin{equation}
	\frac{\hat{\d}}{\hat{\mu}} = \frac{Z_\d}{Z_\mu}\frac{\d_0}{\mu_0}\, ,
\end{equation}
and at or near maximal twist, to a very good approximation, $m_\pi^2 \simeq \hat{\mu}$.  If one further approximate $Z_\d \simeq Z_\mu$, then a conservative estimate can be made for the isospin breaking mass parameter
\begin{equation}
	\hat{\d} \simeq \frac{\d_0}{\mu_0} m_\pi^2\, ,
\end{equation}
where $\d_0$ and $\mu_0$ are the bare mass parameters of the action, Eq.~\eqref{eq:myAction}.  A determination of the physical value of this parameter would allow a precise determination (supplemented by the approximations we have made) of the $up$-$down$ quark mass splitting,
\begin{equation}
	\frac{\hat{\d}}{\hat{\mu}} = \frac{m_d - m_u}{m_d + m_u}\left\{ 1 + \mc{O}\left( \frac{m_\pi^2}{(4\pi f_\pi)^2} \right) \right\}\, .
\end{equation}

\acknowledgments
I would like to thank Kostas Orginos and Steve Sharpe for useful discussions on the subject.  The work of AWL was supported in by the U.S. DOE OJI grant DE-FG02-07ER41527.

\bibliography{tm}

\end{document}